\documentclass[twocolumn,showpacs,preprintnumbers,amsmath,amssymb,superscriptaddress,pra,floatfix]{revtex4}
\usepackage{dcolumn}
\usepackage{bm,bbm}
\usepackage{graphicx,times}
\newcommand{\omegaz}{\omega_{0}}

\newcommand{\ket}[1]{\displaystyle{|#1\rangle}}
\newcommand{\bra}[1]{\displaystyle{\langle #1|}}
\begin{document}
\title{Non-classical correlations in non-Markovian continuous variable systems}
\author{Ruggero Vasile}
\email{ruggero.vasile@utu.fi} \homepage[]{www.openq.fi}
\affiliation{Turku Center for Quantum Physics, Department of Physics
and Astronomy, University of Turku, 20014 Turun Yliopisto, Finland}
\author{Paolo Giorda}
\affiliation{ISI Foundation, I-10133 Torino, Italy}
\author{Stefano Olivares}
\affiliation{CNISM UdR Milano Universit\`a, I-20133 Milano, Italy}
\affiliation{Dipartimento di Fisica, Universit\`a degli Studi di
Milano, I-20133 Milano, Italy}
\author{Matteo G.~A.~Paris}
\affiliation{Dipartimento di Fisica, Universit\`a degli Studi di
Milano, I-20133 Milano, Italy}
\author{Sabrina Maniscalco}
\affiliation{Turku Center for Quantum Physics, Department of Physics
and Astronomy, University of Turku, 20014 Turun Yliopisto, Finland}
\date{\today}
\begin{abstract}
We consider two identical and non-interacting harmonic oscillators
coupled to either two independent bosonic baths or to a common
bosonic bath. Under the only assumption of weak coupling, we analyze
in details the non-Markovian short time-scale evolution of intensity
correlations, entanglement and quantum discord for initial two-mode
squeezed-thermal vacuum states. In the independent reservoirs case
we observe the detrimental effect of the environment for all these
quantities and we establish a hierarchy for their robustness against
the environmental noise. In the common reservoir case, for initial
uncorrelated states, we find that only quantum discord can be
created via interaction with the bath, while entanglement and sub
shot noise intensity correlations remain absent.
\end{abstract}
\pacs{03.67.Mn,03.65.Yz}
\maketitle
\section{Introduction}
Quantum correlations have been the subject of intensive studies in
the last two decades, mainly due to the general belief that they are
a fundamental resource for quantum information processing tasks.
Perhaps, the first rigorous attempt to address the classification of
quantum correlation from an information viewpoint has been put
forward by Werner \cite{Wer}, who introduced an operational
definition of  quantum entanglement as the property of states that
cannot be prepared by local operations  and classical communication
between the two parties. One might have thought that such classical
information exchange could not bring any quantum character to the
correlations in the state. In this sense separability has often been
regarded as a synonymous of classicality of correlations.  However,
it has been shown recently \cite{OllZur,HenVed} that this is not the
case, and a measure of correlations -- quantum discord -- has been
introduced as the mismatch between two quantum analogues of
classically equivalent expressions of the mutual information. For
pure entangled states quantum discord coincides with the entropy of
entanglement. However, quantum discord can be different from zero
also for some (mixed) separable state. In other words, classical
communication can give rise to quantum correlations due to the
existence of non orthogonal quantum states. Quantum discord,
therefore, captures quantum correlations more general than
entanglement. Separable mixed states having nonzero quantum discord
have been proven to provide computational speed up in some quantum
algorithms \cite{DatCav,LanWhi} compared to their classical
counterpart. In addition, the vanishing of quantum discord between
two systems has been shown to be a requirement for the complete
positivity of the reduced subsystem dynamics \cite{Lid09}.

The definition of quantum discord involves an optimization problem
that, in general, can be tackled only for very simple systems. Even
in the simplest bipartite system, i.e., a system of two qubits, an
analytic expression for the discord for the most general two-qubits
state does not exist. The optimization problem has been indeed only
recently solved for the subset of states  which are unitary locally
equivalent to the so-called X-states \cite{Luo,Ali} but remains
unsolved for more general states. For this reason the dynamics of
the discord in presence of the environment has been up to now
studied only for very simple finite-dimensional systems. In the case
of two qubits, e.g., both the Markovian and the non-Markovian time
evolution of quantum correlations have been investigated
\cite{MarDisc,NMDisc1,NMDisc2,NMDisc3} and it has been shown that
discord and entanglement behave differently under the effect of the
environment. In particular, the phenomenon of entanglement sudden
death \cite{YuEbe}, i.e., the complete loss of entanglement after a
finite time, does not occur for quantum discord, which instead
disappears only asymptotically \cite{MarDisc}. Remarkably, for
two-qubits systems in presence of nondissipative noise, the discord
may remain constant in time for very long time intervals, providing
the first evidence of a quantum property that is completely
unaffected by the environmental noise for very long times
\cite{Laura}. The sudden transition from classical to quantum
decoherence, associated to the constant discord phenomenon has been
recently observed experimentally in a quantum optical set up
\cite{NatComm}.

In this paper we present a detailed analysis of the time evolution
of quantum discord for a more involved bipartite system, namely a
system consisting of two non interacting harmonic oscillators
initially prepared in a thermal twin beam (TWB) state. The analytic
formula for the quantum discord for generic bimodal Gaussian states
has been discovered only very recently \cite{Matt,DaAde}. We use
such definition to evaluate how the quantum correlations evolve in
presence of both independent and common bosonic thermal reservoirs.

The system we are going to analyze have an immediate application in
a quantum optical setting where it may be implemented by parametric
downconversion (PDC), which has been addressed as a convenient and
feasible setting to visualize the evolution of quantum correlations
\cite{DBA07,ss07}. In turn, the pair of field modes obtained from
thermally seeded PDC is as a convenient physical system to analyze
the quantum-classical transition in the continuous variable regime.
This scheme have been already investigated in ghost-
imaging/diffraction experiments \cite{DBA07}, where it has been
shown that both entanglement and intensity correlations may be tuned
upon changing the intensities of the seeds \cite{DBA07,ss07}.
In this framework, besides fundamental quantities like entanglement
and quantum discord, we will also evaluate a more operational
quantity as the degree of correlations between the intensities of
the two beams exiting the noisy channel. The shot-noise limit (SNL)
in a photodetection process is defined as the lowest level of noise
that can be achieved by using semiclassical states of light
\cite{SNL}, that is Glauber coherent states. On the other hand, when
a noise level below the SNL is observed, we have a genuine
nonclassical effect. For a two-mode system if one measures the
photon number of the two beams and evaluates the difference
photocurrent the SNL is the lower bound to the fluctuations that is
achievable with classically coherent beams and a noise level below
the SNL indicates the presence of nonclassical correlations between
the beams.

We consider the case in which system and environment are weakly
coupled but we do not perform the Markov approximation so our
results also described the initial short time correlations between
the system and the reservoir. The lifetime of such correlations
depends on the structure of the environment. When the spectral
density of the environment changes significantly for frequencies
close to the system characteristic frequency, the correlations
between system and reservoir persist for a longer time and
non-Markovian approaches are necessary.

The dynamics of entanglement in such structured reservoirs has been
studied in both the common \cite{PraBer,PazRon} and the independent
reservoir scenario \cite{PraBer,IndepRes,VasMan}. Here we compare
the time evolution of the discord with the one of both the
entanglement and the intensity correlations. In the case of
independent reservoirs, we can establish a hierarchy of
nonclassicality markers in terms of their robustness against the
destructive action of the environment. In the common reservoir
scenario we find that if the initial state does not possess  quantum
correlations, i.e., all three markers of nonclassicality here
considered have initially zero value, as time passes the interaction
with the common reservoir can create quantum discord between the two
system oscillators. Entanglement and nonclassical intensity
correlations, however, cannot be created by the common reservoir in
the weak coupling limit here studied. Finally, we analyze how the
quantum discord behaves as a function of the initial thermal
component of the TWB state. We discover that this quantity
influences the rate of change of the discord in both the independent
and the common reservoir cases.

The paper is structured as follows. In Sec.~\ref{s:model} we present
the microscopic physical models for the system and the reservoir. In
Sec.~\ref{s:markers} we introduce the three markers of
nonclassicality and, in particular, of nonclassical correlations
considered in this paper: intensity correlations, entanglement and
discord. Sec.~\ref{s:dynam} investigates the dynamics of the markers
in presence of common or independent reservoirs. Finally,
Sec.~\ref{s:concl} closes the paper and draws some concluding remark.

\section{Physical models}\label{s:model}
In this section we introduce a physical model widely used in the
description of the non-Markovian dynamics of CV quantum channels.
The main system is made of a pair of identical non-interacting
harmonic oscillators of frequency $\omegaz$ and unit mass. The free
Hamiltonian reads
\begin{equation}
\hat{H}_0= \hat H^0_1 + \hat H^0_2
=\frac{1}{2}\sum_{j=1,2}(\hat{P}_j^2+\omegaz^2 \hat{X}_j^2),
\end{equation}
where $\hat{P}_j=\frac{1}{i\sqrt{2}}(\hat{a}_j-\hat{a}_j^\dag)$ and
$\hat{X}_j=\frac{1}{\sqrt{2}}(\hat{a}_j+\hat{a}_j^\dag)$ are the
momentum and position operators, respectively, and $\hat{a}_j$ the
field operator of the harmonic oscillators (the index $j=1,2$ labels
the oscillators). Additionally, we suppose that the harmonic
oscillators interact with an external environment. In the following
we introduce two different interaction models.

\subsection{Independent reservoirs}

The first model consists of an external environment made of two
independent bosonic baths with free Hamiltonian
\begin{equation}
\hat{H}_B=\sum_{j,k}\biggl(\frac{\hat{\Pi}_{jk}^2}{2m_{jk}}+\frac{m_{jk}
w_{jk}^2\hat{Q}_{jk}^2}{2}\biggl).
\end{equation}
The index $j=1,2$ labels the bath, and $k$ runs over all the bath
modes. The $\hat{\Pi}_{jk}$ ($\hat{Q}_{jk}$) are the momentum
(position) operators, while $w_{jk}$ and $m_{jk}$ are the
frequencies and masses associated to each bosonic mode.

Each system oscillator interacts with its own bosonic bath (same
index $j$) through a position-position coupling described by the
following interaction Hamiltonian
\begin{equation}\begin{split}
&\hat{H}_I=\alpha\sum_{j,k}\gamma_{jk}\hat{X}_j\hat{Q}_{jk},
\end{split}\end{equation}
where $\gamma_{jk}$ are the coupling constants between the $j$-th
oscillator and the $k$-th mode of its bath and $\alpha$ is a
dimensionless coupling constant. For the sake of simplicity hitherto
we assume that the baths have the same spectral structure and are
equally coupled to the oscillators.

The reduced dynamics of the two oscillators in the case of
stationary reservoirs is described by the following exact time-local
master equation \cite{HuPaZa}
\begin{align}
\dot{\varrho}(t)=&\sum_j\frac{1}{i\hbar}[\hat{H}_j^{0},\varrho(t)]-\Delta(t)
[\hat{X}_j,[\hat{X}_j,\varrho(t)]]\nonumber \\
&+\Pi(t)[\hat{X}_j,[\hat{P}_j,\varrho(t)]]+
\frac{i}{2}r(t)[\hat{X}_j^2,\varrho(t)]\nonumber \\
&-i\gamma(t)[\hat{X}_j,\{\hat{P}_j,\varrho(t)\}],\label{HuPaZang}
\end{align}
where $\varrho(t)$ is the reduced density operator of the
oscillators and $\hat{H}_j^{0}$ is the free Hamiltonian of the
$j$-th oscillator. The time dependent coefficients, describing
diffusion ($\Delta(t)$, $\Pi(t)$), damping ($\gamma(t)$) and free
frequency renormalization($r(t)$) processes, can be expressed as
power series in the system-reservoir coupling constant $\alpha$. In
the weak coupling limit we can stop the expansion to second order in
$\alpha$ and obtain analytic solutions for the coefficients. We
provide their expressions in the Appendix for high temperature
reservoirs characterized by an Ohmic spectral density with
Lorentz-Drude cutoff
$J(\omega)=\frac{\omega_c^2}{\pi}\frac{\omega}{\omega^2+\omega_c^2}$.

Using the characteristic function approach the solution of
\eqref{HuPaZang} in the weak coupling limit is given by \cite{Intra}
\begin{align}
\chi_t(\Lambda)=&\mbox{} \exp\{-\Lambda^T
[\bar{\mathbf{W}}(t)\oplus\bar{\mathbf{W}}(t)]\Lambda\} \nonumber \\
&\times\chi_0(e^{-\Gamma(t)/2}[\mathbf{R}^{-1}(t)\oplus
\mathbf{R}^{-1}(t)]\Lambda), \label{CharSolInd}
\end{align}
where $\chi_t(\Lambda)$ is the characteristic function at time $t$,
$\chi_0$ is the characteristic function at the initial time $t=0$,
$\Lambda = (x_1,p_1,x_2,p_2)^T$ is the two-dimensional phase space
variables vector, $\Gamma(t)=2\int_0^t\gamma(t')dt'$, and
$\bar{\mathbf{W}}(t)$ and $\mathbf{R}(t)$ are $2 \times 2$ matrices
whose expression is also given in the Appendix.

The interaction between oscillators and baths is bilinear in
position and momentum, thus it induces a Gaussian evolution. This is
of great importance because, as we will see, analytic expressions
for quantum correlations can be obtained only in the case of
Gaussian states.

The characteristic function of a Gaussian state with zero mean
depends only on the expression of the covariance matrix $\sigma$,
whose elements are defined as
$\sigma_{ij}=\langle\{\hat{\Lambda}_i,\hat{\Lambda}_j\}\rangle/2-\langle\hat{\Lambda}_i\rangle\langle\hat{\Lambda}_j\rangle$,
where $\hat{\Lambda}\equiv(\hat{X}_1,\hat{P}_1,\hat{X}_2,\hat{P}_2)$
and $\langle\cdot\rangle$ indicates the mean value over the state,
\begin{equation}\label{Charfun}
\chi_0(\Lambda)=\exp\biggl\{-\frac{1}{2}\Lambda^T\sigma(0)\Lambda\biggl\}.
\end{equation}
Using Eqs. \eqref{CharSolInd} and \eqref{Charfun} we get the
evolution of the covariance matrix under the action of the two
independent reservoirs
\begin{equation}\begin{split}\label{CovMatInd}
\sigma(t)=&e^{-\Gamma(t)}[\mathbf{R}(t)\oplus
\mathbf{R}(t)]\sigma(0)[\mathbf{R}(t)\oplus
\mathbf{R}(t)]^{-1}\\
&+2[\bar{\mathbf{W}}(t)\oplus\bar{\mathbf{W}}(t)].
\end{split}\end{equation} The solution in the case of an initial symmetric
covariance matrix in its normal form is
\begin{equation}\label{CovMat}
\sigma(0)=\left(
         \begin{array}{cc}
           \mathbf{A_0} & \mathbf{C_0} \\
           \mathbf{C_0} & \mathbf{A_0} \\
         \end{array}
       \right)\Rightarrow\sigma(t)=\left(
         \begin{array}{cc}
           \mathbf{A^{ind}_t} & \mathbf{C^{ind}_t} \\
           \mathbf{C^{ind}_t} & \mathbf{A^{ind}_t} \\
         \end{array}
       \right),
\end{equation}
with $\mathbf{A_0}= a\,{\mathbbm 1}$, $\mathbf{C_0}={\rm
  diag}(c_1,c_2)$, $a>0$ and $c_1$, $c_2$ real numbers, and $\mathbbm
1$ the $2\times 2$ identity matrix (note that $\mathbf{C_0}$ and $
\mathbf{C^{ind}_t}$ are symmetric matrices). The analytic expression
of the matrices $\mathbf{A^{ind}_t}$ and $\mathbf{C^{ind}_t}$ is
given in the Appendix.

\subsection{Common reservoir}

In the second example of system-environment interaction model, we
consider a common bosonic bath, and look at the case in which both
system oscillators interact with it symmetrically. The Hamiltonian
reads
\begin{equation}\begin{split}
&\hat{H}_B+\hat{H}_I=\sum_k\biggl(\frac{\hat{\Pi}_k^2}{2m_k}+\frac{m_k
w_k^2\hat{Q}_k^2}{2}\biggl)+\alpha\sum_{j,k}\gamma_k\hat{X}_j
\hat{Q}_k.
\end{split}\end{equation}
In order to write the reduced dynamics solution $\varrho(t)$, we
first apply a canonical transformation to the Hamiltonian following
the lines of \cite{PazRon}. We define new position
$\hat{X}_{\pm}=(\hat{X}_1\pm \hat{X}_2)/\sqrt{2}$ and momentum
$\hat{P}_{\pm}=(\hat{P}_1\pm \hat{P}_2)/\sqrt{2}$ operators for the
system. Under this transformation the total Hamiltonian becomes
\begin{subequations}
\begin{align}
H_0&=
\frac{\hat{P}_+^2+\hat{P}_-^2}{2}+\frac{\omega_0^2}{2}(\hat{X}_+^2+\hat{X}_-^2)\\
H_R&=\sum_k\biggl(\frac{\hat{\Pi}_k^2}{2m_k}+\frac{m_k
w_k^2\hat{Q}_k^2}{2}\biggl)\\
H_I&=\alpha\sqrt{2}\hat{X}_+\sum_k \gamma_k \hat{Q}_k
\end{align}
\end{subequations}
In this picture only one oscillator interacts with the bath through
a position-position coupling, the other evolving freely. It follows
that the master equation for the reduced state $\tilde{\varrho}(t)$
in the new picture becomes
\begin{align}
\dot{\tilde{\varrho}}(t)=&
\frac{1}{i\hbar}[(\hat{H}_-^{0}+\hat{H}_+^{0}),\tilde{\varrho}(t)]
-\sqrt{2}\Delta(t) [\hat{X}_+,[\hat{X}_+,\tilde{\varrho}(t)]]\nonumber\\
&+\sqrt{2}\Pi(t)[\hat{X}_+,[\hat{P}_+,\tilde{\varrho}(t)]]
+\frac{i}{\sqrt{2}}r(t)[\hat{X}_+^2,\tilde{\varrho}(t)]\nonumber\\
&-i\sqrt{2}\gamma(t)[\hat{X}_+,\{\hat{P}_+,\tilde{\varrho}(t)\}],\label{HuPaZangCom}
\end{align}
This is of the same form of \eqref{HuPaZang}, except for the fact
that only one effective oscillator is coupled to the environment.
The dynamics in terms of the characteristic function is then
\begin{align}
\tilde{\chi}_t(\Lambda_{\pm})=&\exp\{-\Lambda^T_{\pm}
[\sqrt{2}\bar{\mathbf{W}}(t)\oplus\mathbf{0}]\Lambda_{\pm}\}\nonumber \\
&\times\tilde{\chi}_0([e^{-\frac{\Gamma(t)}{\sqrt{2}}}\mathbf{R}^{-1}(t)\oplus
\mathbf{R}^{-1}(t)]\Lambda_{\pm}), \label{CharSolCom}\end{align}
with $\Lambda_{\pm}=(x_+,p_+,x_-,p_-)^T$ and $\mathbf{0}$ being the
$2\times 2$ zero matrix. Equivalently, the associated covariance
matrix $\tilde{\sigma}(t)$ evolves as
\begin{align}
\tilde{\sigma}(t)=&\left[e^{-\frac{\Gamma(t)}{\sqrt{2}}}\mathbf{R}(t)\oplus
\mathbf{R}(t)\right] \tilde{\sigma}(0)
\left[e^{-\frac{\Gamma(t)}{\sqrt{2}}}\mathbf{R}(t)\oplus \mathbf{R}(t)\right]^{-1}\nonumber\\
&+2\sqrt{2}[\bar{\mathbf{W} }(t)\oplus\mathbf{0}].\label{CovMatComm}
\end{align}
As in the previous case an initial Gaussian state will maintain its
character during the time evolution. Indeed, the canonical
transformation, its inverse and the dynamical evolution are all
Gaussian operations.

Given the initial covariance matrix $\sigma(0)$, applying the
transformations and using Eqs. \eqref{CharSolCom} and
\eqref{CovMatComm}  we get
\begin{equation}\label{CovMat:com}
\sigma(0)=\left(
         \begin{array}{cc}
           \mathbf{A_0} & \mathbf{C_0} \\
           \mathbf{C_0} & \mathbf{A_0} \\
         \end{array}
       \right)\Rightarrow\sigma(t)=\left(
         \begin{array}{cc}
           \mathbf{A^{com}_t} & \mathbf{C^{com}_t} \\
           \mathbf{C^{com}_t} & \mathbf{A^{com}_t} \\
         \end{array}
       \right),
\end{equation}
with $ \mathbf{C^{com}_t}$ symmetric matrix. Details of the solution
are still given in the Appendix.

\section{Non-classical Correlations}\label{s:markers}

In the last decades there has been a growing interest in the issue
of identifying  and possibly quantifying the quantumness of states
of a given physical system. One of the reasons is that states
possessing quantum features may be useful for certain quantum
information and computation protocols, or in the field of precision
measurements,  enhancing computation and measurements efficiencies.

In the case of bipartite (or in general multipartite) systems the
interest is directed not only towards the quantumness of the state
itself, but also towards the quantumness of correlations between the
different parts. In this paper we provide new insight on this issue
by comparing different markers of quantumness of states and
correlations. In particular we are interested in studying how
non-Markovian dynamical evolutions affect these quantities in the
context of two mode continuous variable systems. In the following we
introduce the well-known concepts of intensity correlations and
entanglement in CV systems as well as the recently introduced
quantum discord for Gaussian states.

\subsection{Intensity correlations}

Firstly we consider the intensity correlations marker
$\mathbf{I}_{\rm corr}$, which is related to the measurement of the
two light beams intensities, i.e., $\langle\hat{n}_1\rangle$ and
$\langle\hat{n}_2\rangle$, with
$\hat{n}_i=(\hat{X}_i^2+\hat{P}_i^2)/2$ being the number operator of
the $i$-th mode, and thus feasible with current technology. More
precisely, the intensity correlations marker is defined as
\cite{DBA07,Icorr:brida}
\begin{equation}\label{Icorr}
\mathbf{I}_{\rm corr}=1-
\frac{\langle\Delta\hat{I}_-^2\rangle}{\langle\hat{n}_1
+\hat{n}_2\rangle},
\end{equation}
and is based on the measurement of the operator
$\hat{I}_{-}=\hat{n}_1-\hat{n}_2$, that is the difference between
the intensities of the two light modes, whose variance
$\langle\Delta \hat{I}_-^2\rangle$ may also be written as
\begin{align}
\langle\Delta \hat{I}_-^2\rangle &= \langle \hat{I}_-^2\rangle-\langle
\hat{I}_-\rangle^2,\\
&= \langle \Delta \hat n_{1}^2 \rangle +
\langle \Delta \hat n_{2}^2 \rangle - 2
\langle \hat n_{1} \rangle \langle \hat n_{2} \rangle\,
g^{(2)}(\hat n_{1},\hat n_{2}),
\end{align}
where we introduced the second-order correlation function
\begin{equation}
g^{(2)}(\hat n_{1},\hat n_{2}) =
\frac{\langle \hat n_{1} \hat n_{2} \rangle}
{\langle \hat n_{1} \rangle \langle \hat n_{2} \rangle} -1.
\end{equation}
In the case of products of coherent states we have $\mathbf{I}_{\rm
corr}=0$, which defines the shot-noise limit (SNL) for this
particular detection process, that is the lowest level of noise that
can be obtained by using the semiclassical states of light, i.e.,
the coherent states. On the other hand, when
\begin{equation}
0<\mathbf{I}_{\rm corr}\leq1,
\end{equation}
the fluctuations on the intensity correlations are below the SNL,
indicating genuine non-classical features in the state of the system.
It is worth stressing that intensity correlations below the SNL can be
observed also for product states, for example in the presence of local
squeezing. Hence, this feature is not necessary related to the
quantumness of correlations among different parts of our bipartite
system, but rather to the quantumness of the overall state itself
\cite{ferr:prep}.

For a Gaussian state with zero mean value, $\mathbf{I}_{\rm corr}$
depends only on the corresponding covariance matrix $\sigma$ and, in
the case of symmetric Gaussian states, reads
\begin{equation}
\mathbf{I}_{\rm corr}=1-
\frac{\sigma^2_{11}+\sigma^2_{22}+2\sigma_{13}^2-\sigma_{14}^2-\sigma_{23}^2-\sigma_{24}^2-\frac{1}{2}}{\sigma_{11}+\sigma_{22}-1},
\end{equation}
where $\sigma_{ij}$ are the covariance matrix entries.

\subsection{Entanglement}

Entanglement dynamics in dissipative bipartite continuous variable
domain has been object of interest and numerous studies in recent
years \cite{PraBer,PazRon,IndepRes,VasMan}. Though there exist
separability criteria and entanglement measures for a bipartite
Gaussian state $\varrho$ (see, e.g.,
\cite{Simon,EntFor,buono:JOSAB:10}), in this paper we study the
entanglement dynamics by focusing on the logarithmic negativity
defined as \cite{Nega}
\begin{equation}
\mathcal{N}(\varrho)=\max\{0,-\log(2\tilde{\nu}_-)\},
\end{equation}
with $\tilde{\nu}_-$ being the minimum symplectic eigenvalue of the
partially transpose ($\mathbf{PT}$) covariance matrix of the system,
namely, $\sigma^{\rm PT} = {\boldsymbol \Delta} \sigma {\boldsymbol
  \Delta}$ with $ {\boldsymbol \Delta}=\hbox{diag}(1,-1,1,-1)$. It is
worth to note that $\mathcal{N}(\varrho)>0$ if and only if
$\tilde{\nu}_{-}<1/2$, that is if and only if the state $\varrho$ is
entangled: of course, the condition $\tilde{\nu}_{-}<1/2$ is a necessary
and sufficient condition for a bipartite Gaussian state to be non
separable \cite{Simon}.

\subsection{Quantum Discord}

The total amount of correlations in a bipartite quantum system
having density operator $\varrho$ is quantified by the quantum
version of the mutual information
\begin{equation}
\mathcal{I}(\varrho)=S(\varrho_1)+S(\varrho_2)-S(\varrho),
\end{equation}
where $S(\cdot)$ is the Von Neumann entropy, and
$\varrho_{1(2)}=\hbox{Tr}_{2(1)}[\varrho]$. Usually the total correlations
are divided in a quantum part, known as \emph{quantum discord}
$\mathcal{D}(\varrho)$, and a classical part
$\mathcal{C}(\varrho)$. The classical correlations are defined as
the maximum amount of information we can gain on one part of the
system by locally measuring the other subsystem \cite{HenVed},
\begin{equation}\begin{split}
\mathcal{C}(\varrho)= \max_{\Pi_i}\bigl\{S(\varrho_1)-\sum_i p_i
S(\varrho^{\Pi_i}_{1|2})\bigl\},
\end{split}
\label{ClassCorrs}\end{equation} where
$\varrho^{\Pi_i}_{1|2}=\mbox{Tr}_2(\varrho \openone\otimes\Pi_i)$ is
the post measurement state in which system $1$ is left when the
result $i$ occurs in a measurement of system $2$ with probability
$p_i=\mbox{Tr}_{1\, 2}(\varrho \openone\otimes\Pi_i)$. The maximum
is taken over the all positive operator valued measures $\{ \Pi_i
\}$ (POVM), $\sum_i \Pi_i =\openone$ performable on one subsystem.
Classical correlations are thus obtained in correspondence of the
POVM that minimizes the conditional entropy $\sum_i p_i
S(\varrho^{\Pi_i}_{1|2})$ i.e., that allows one to obtain the
highest amount of information on the state of system $1$. The above
definition is in general non symmetric with respect to the
interchange of the subsystems. In our case however, due to our
specific choice of the system's initial states (see below) and to
the symmetry of the coupling with the bath, (\ref{ClassCorrs}) turns
out to be symmetric during the entire evolution; therefore no
specific indication of the subsystem measured will be needed.  The
quantum discord is then defined as the difference between the total
correlations and the classical correlations
\begin{equation}\begin{split}
&\mathcal{D}(\varrho)=\mathcal{I}(\varrho)-\mathcal{C}(\varrho).
\end{split}\end{equation}
A peculiar property of quantum discord is that it can be nonzero
even if the state is separable. This is an indication of the fact
that entanglement is not the only source of quantum correlations.
Recently examples of quantum computational algorithms showing a
speedup with respect to the classical counterparts, also in the
absence of entanglement, have been presented \cite{DatCav,LanWhi}.
It is believed that the presence of quantum correlations
other-than-entanglement is responsible for this feature. In this
sense it is important to study how the quantum discord evolves in
presence of the external environments, comparing, e.g., its behavior
with the behavior of entanglement. In the following we answer to
this question for Gaussian states of CV systems.

To evaluate the total quantum correlations we use a recently
developed expression valid only for Gaussian states
\cite{Matt,DaAde}. Given the block form of the covariance matrix
$\sigma(t)$ \eqref{CovMat} and \eqref{CovMat:com}, in the symmetric
case, the Gaussian quantum discord is defined as
\begin{equation}
\mathcal{D}(\varrho)=f\left(\sqrt{{\rm det}\mathbf{A_t}}\right)+
f\left(\sqrt{{\rm det}\mathbf{\tau}}\right)-f(n_+)-f(n_-),
\end{equation}
where
$f(x)=(x+\frac{1}{2})\ln(x+\frac{1}{2})-(x-\frac{1}{2})\ln(x-\frac{1}{2})$,
$n_{\pm}$ are the symplectic eigenvalues of the covariance matrix
and
\begin{equation}
\tau=\mathbf{A_t}-\mathbf{C_t}(\mathbf{A_t}+\sigma_M)^{-1}\mathbf{C_t}^T,
\end{equation}
is the covariance matrix of the state of the system $A$ after the
generalized Gaussian measurement on the system $B$, described by the
covariance matrix
\begin{equation}
\sigma_M=\frac{\cosh 2\rho}{2}\left(
          \begin{array}{cc}
            1+\tanh 2\rho\cos\phi & -\tanh 2\rho\sin\phi \\
            -\tanh 2\rho\sin\phi & 1-\tanh 2\rho\cos\phi \\
          \end{array}
        \right),
\end{equation}
with $\rho\geq0$ and $0\leq\phi\leq 2\pi$. For a generic Gaussian
state we must perform a minimization procedure in order to find the
appropriate generalized measurement for the calculus of the quantum
discord. In the case of $\mathbf{C_t}={\rm diag}(c,-c)$ the minimum
is obtained for a completely heterodyne measurement \cite{Matt}. For
a generic covariance matrix in its normal form the exact expression
for the discord has been evaluated in \cite{DaAde}. However in our
cases the time evolution of the covariance matrix is not in the
normal form. Therefore to evaluate the discord it was faster to
implement a numerical minimization procedure.

\section{Results}\label{s:dynam}

In this section we study the evolution of the previously introduced
markers of nonclassicality under the influence of either independent
or common reservoirs, in the weak coupling regime. We limit our
investigation to the case of an Ohmic spectrum with Lorentz-Drude
cut-off at high temperatures focusing on the non-Markovian short
time-scale. Moreover we fix hitherto the coupling constant
$\alpha=0.1$ and $k_BT/\hbar\omega_c=100$, according respectively to
the weak coupling and high temperature assumptions.

An important parameter in our discussion is the ratio between the
cut-off frequency of the baths spectrum $\omega_c$ and the
oscillator frequency $\omega_0$, namely the \emph{resonance
parameter} $x=\omega_c/\omega_0$. In \cite{VasMan} we have studied
the non-Markovian entanglement dynamics noting the existence of two
dynamical regimes ($x\ll 1$ and $x\gg 1$) characterized by
qualitatively and quantitatively different dynamical behaviors.
Genuine non-Markovian effects occur in the $x\ll 1$ regime because
the time-dependent coefficients in the Master equation attain
negative values in certain time intervals. This feature leads to
entanglement oscillations, which are not present in the $x\gg 1$
case. As we will see, the same conclusion is valid for the intensity
correlations and the quantum discord, also in the common reservoir
scenario. Moreover, in general, the dynamics for $x\ll 1$ is much
slower. In the following we concentrate especially on the $x\gg 1$
regime, or linear spectrum regime, unless qualitatively different
phenomena can be reported in the other regime.

Let us consider as initial states the thermal TWB states defined as
\begin{equation}
\varrho_{in}(r,N_1,N_2)=
\hat{S}_2(r)\nu(N_1)\otimes\nu(N_2)\hat{S}_2^{\dag}(r),
\end{equation}
where $\nu(N)=\sum_n N^n(1+N)^{-(n+1)}\ket{n}\bra{n}$ is a thermal
state with $N$ average photons and
$\hat{S}_2(r)=\exp\{r(\hat{a}_1^{\dag}\hat{a}_2^{\dag}-\hat{a}_1\hat{a}_2)\}$
is the two-mode squeezing operator. In the symmetric case we are
interested in the two thermal states are characterized by the same
temperature parameter $N_1=N_2=N$, while the squeezing parameter $r$
can assume any non negative value. The covariance matrix $\sigma(0)$
of the initial state is given by
\begin{equation}
\mathbf{A}_0=\left(
                                        \begin{array}{cc}
                                          a & 0 \\
                                          0 & a \\
                                        \end{array}
                                      \right),\qquad
\mathbf{C}_0=\left(
                                        \begin{array}{cc}
                                          c & 0 \\
                                          0 & -c \\
                                        \end{array}
                                      \right),
\end{equation}
with $a=(N+1/2)\cosh(2r)$ and $c=(N+1/2)\sinh(2r)$. If $r=0$ the
state is initially uncorrelated. If $r>0$ the state may be entangled
or separable, depending on the value of $N$,  but it will always
possess non-zero quantum discord \cite{Matt}.

\subsection{Independent reservoirs}

Initially uncorrelated states cannot become correlated at a later
time when evolving under local operations, as in the case of the
independent reservoirs model. Thus we focus here on initially
correlated states ($r>0$). In all the various examples we examined,
we observed that the interaction with the reservoirs has a
detrimental effect for all the quantumness markers introduced in the
previous section. In this sense, not only the state becomes more
classical but also the quantum correlations decrease.

Entanglement however behaves differently from quantum discord.
Indeed entanglement can disappear after a finite time, exhibiting a
sudden death and, depending on bath parameters and temperature, also
exhibit partial revivals \cite{VasMan}. On the contrary, in our
system the quantum discord vanishes only for $t\rightarrow\infty$, a
result which is independent from the value of the resonance
parameter and, at least in the weak coupling limit, it is also
independent from the spectral distribution and temperature regime.
This is a consequence of the fact that Gaussian quantum discord is
zero if and only if the Gaussian state is a product state and
therefore if and only if the determinant of the $C$ matrix is zero
\cite{Matt,DaAde}. This condition is never satisfied in weak
coupling case for initial two-mode squeezed thermal states (see
Appendix).

In FIG. \ref{fig:1} we show the behavior of $\mathbf{I}_{\rm corr}$
(only the sub-shot noise regime), $\mathcal{N}(\varrho)$ and
$\mathcal{D}(\varrho)$ for initial states with $r=2$ and $N=0$ as
a function of $\omega_c t$. Note that, in the figures we have scaled
all these quantities so that their initial value coincides. In FIG.
\ref{fig:1} (top) we choose $x=10$, while in FIG. \ref{fig:1}
(bottom) we have $x=0.2$. The dynamics is faster when $x$ is large
and it does not present oscillations typical of a non-Markovian
evolution. Moreover entanglement is more robust to the detrimental
effect of the environment than $\mathbf{I}_{\rm corr}$. This is valid
also for $x\ll 1$, a fact that helps us to set up a hierarchy for
the behavior of our quantities under the influence of the
independent environment. We will see that such a classification
cannot be done instead in the more complicated dynamics due to a
common reservoir.

On the other hand when $x\ll 1$ non-Markovian oscillations are
present in all quantum markers. Moreover, in this case, intensity
correlations and entanglement go to zero at the same time,
independently from the initial thermal squeezed state.

\begin{figure}[h!]
\begin{center}
\includegraphics[width=6cm]{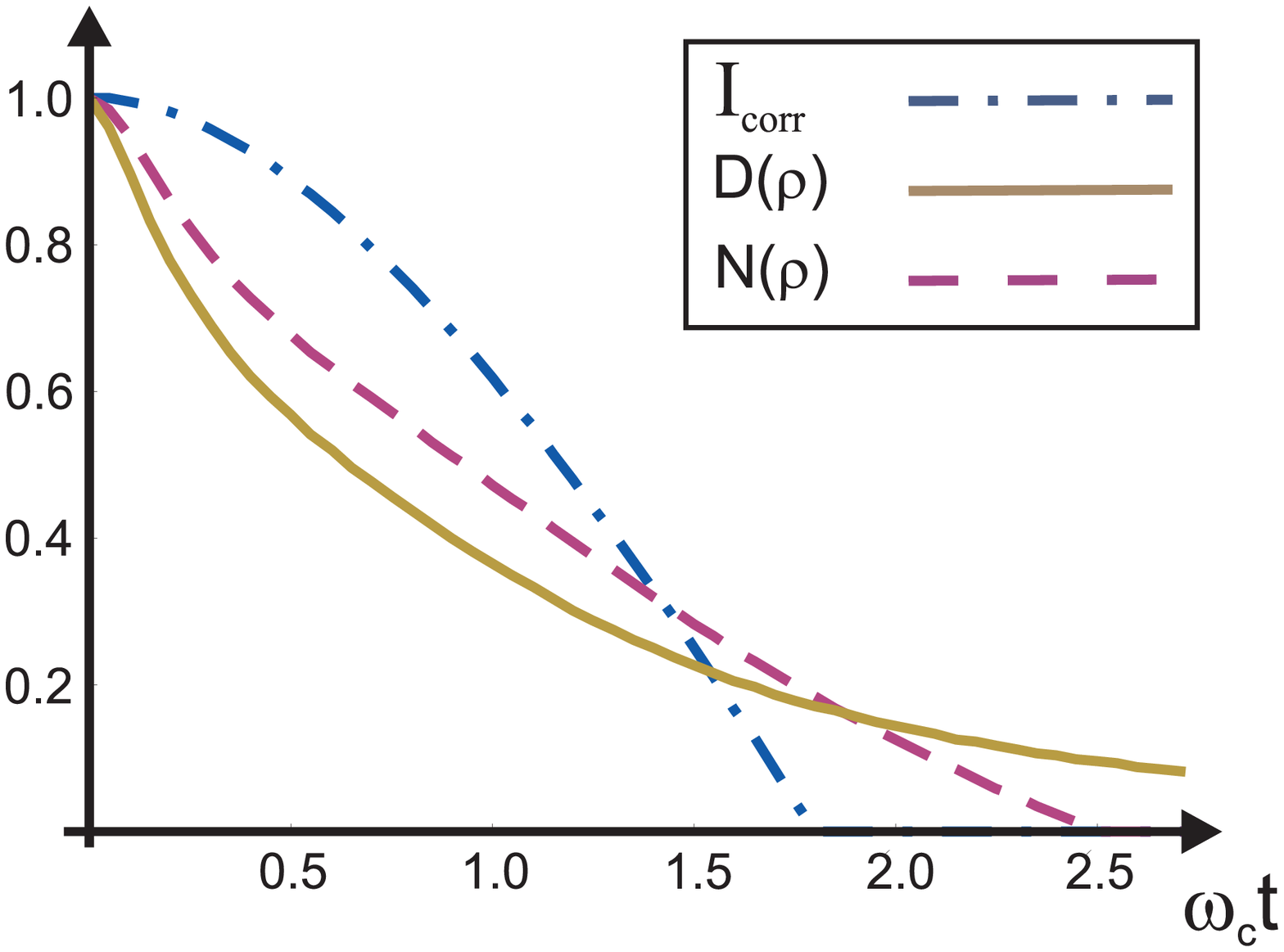}
\includegraphics[width=6cm]{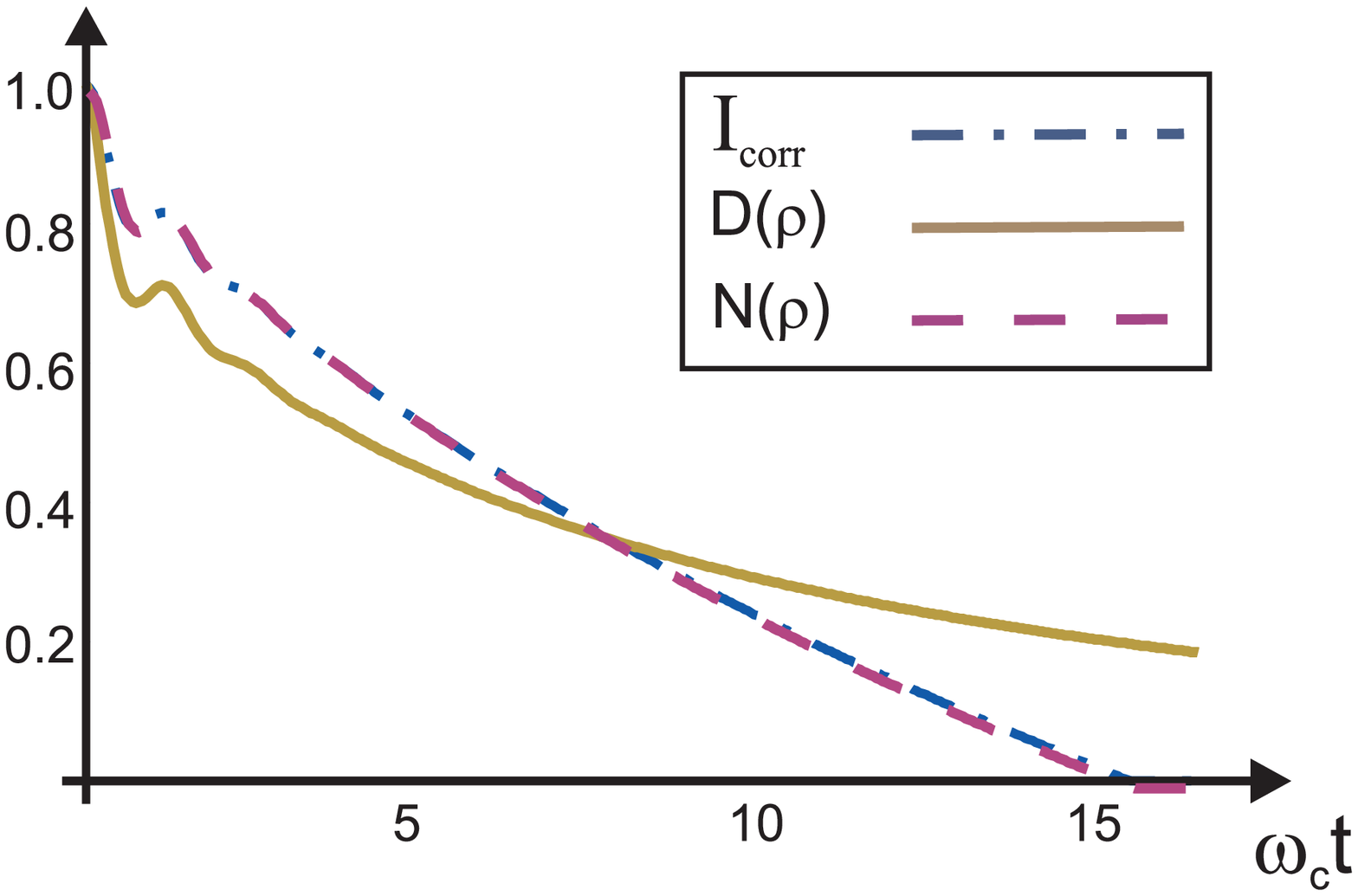}
\end{center}
\vspace{-0.5cm} \caption{(Colors online) Evolution of intensity
correlations marker $\mathbf{I}_{\rm corr}$ below the SNL (blue dashed
line), logarithmic negativity $\mathcal{N}(\varrho)$ (red dashed
line) and quantum discord $\mathcal{D}(\varrho) $ (solid yellow
line) in the independent reservoir case as a function of $\omega_c
t$. Parameters: $\alpha=0.1$, $k_BT/\hbar\omega_c=100$ and (top)
$x=10$, $r=2$, $N=0$; (bottom) $x=0.2$, $r=0.5$, $N=0$.}
\label{fig:1}
\end{figure}

\begin{figure}[h!]
\begin{center}
\includegraphics[width=6cm]{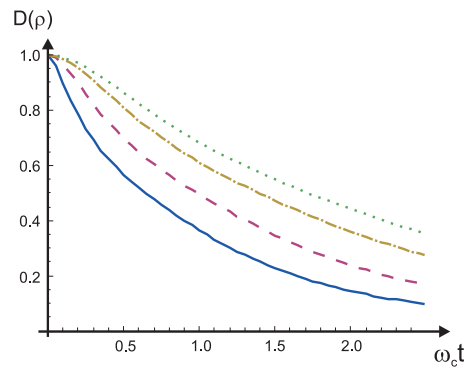}
\end{center}
\vspace{-0.5cm} \caption{(Colors online) Evolution of quantum
discord $\mathcal{D}(\varrho)$ as a function of the scaled time
$\omega_c t$ in the independent reservoir scenario. Parameters:
$\alpha=0.1$, $k_BT/\hbar\omega_c=100$, $x=10$ and $r=2$. Different
lines represent different values of the thermal parameter: $N=0$
(Solid blue line), $N=1$ (Dashed red line), $N=5$ (Dotted-dashed
yellow line) and $N=10$ (Dotted green line).} \label{fig:2}
\end{figure}

We conclude this section with an analysis of the dynamics as a
function of the thermal parameter $N$ of the initial state,
concentrating in particular on the dynamics of the quantum discord.
In FIG. \ref{fig:2} we show the time evolution of
$\mathcal{D}(\varrho)$ for $x=10$, $r=2$ and different values of
$N=0,1,5,10$. One can clearly see that the higher is $N$ the slower
is the loss rate of quantum correlations under the action of local
bosonic baths. In other words, states with initially higher thermal
component loose quantum correlations, as measured by the discord,
more slowly than states with smaller thermal component.

\subsection{Common reservoir}

\begin{figure}[h!]
\begin{center}
\includegraphics[width=6cm]{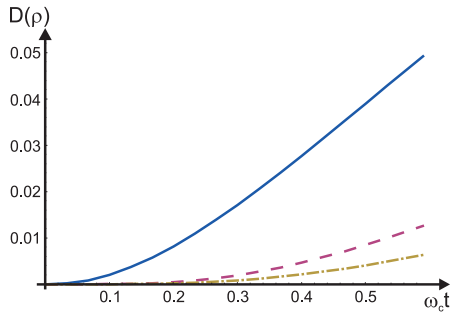}
\end{center}
\vspace{-0.5cm} \caption{(Colors online) Evolution of quantum
discord $\mathcal{D}(\varrho)$ as a function of the scaled time
$\omega_c t$ in the common reservoir scenario. Parameters:
$\alpha=0.1$, $k_BT/\hbar\omega_c=100$, $x=10$ and $r=0$ (initially
uncorrelated state). Different lines represent different values of
the thermal parameter: $N=0$ (Solid blue line), $N=0.05$ (Dashed red
line), $N=0.1$ (Dotted-dashed yellow line).} \label{fig:3}
\end{figure}

In the common reservoir case the CV system dynamics is much richer
than in the independent reservoir case. Initially uncorrelated
states, e.g, become in general correlated as time passes. For the
class of initial Gaussian states considered in the paper, however,
and in the weak coupling limit, we find that, if $r=0$, neither
entanglement nor intensity correlations below the SNL are created
(always $\mathbf{I}_{\rm corr}<0$ ). On the contrary, quantum
discord is created by the action of the common environment and grows
as time passes, for any the value of $N$ and of the resonance
parameter $x$, as shown in FIG. \ref{fig:3} for $x=10$. This result
holds also in the Markovian case \cite{PraBer}. Therefore, in CV
systems, the common reservoir always creates quantum correlations in
the weak coupling regime.

We note that the initial value of $N$ affects the rate of change of
the quantum discord similarly to the independent reservoir scenario
discussed in the previous section. While there, for initial
correlated states, the higher was $N$ the slower was the decrease of
the discord, here, for initially uncorrelated states, the higher is
$N$ the slower is the increase of the discord, as one can see from
FIG. \ref{fig:3}, in the $x=10$ case. This result holds for any
value of the resonance parameter $x$.

We now turn to the initially correlated case, i.e., initial states
having $r>0$ and $N\geq0$. In the common reservoir scenario a
comparison between the dynamics of the three markers, as the one
presented for independent reservoirs, does not provide interesting
information. The reason is that it is not possible to identify a
general hierarchy for the most robust quantities under the action of
the environment. Indeed the dynamics is more strongly dependent on
the initial state and reservoir parameters and the robustness of
each of the markers changes case by case without exhibiting a
general trend. Therefore we will present in the following the most
interesting dynamical features of each quantity separately.

\begin{figure}[h!]
\begin{center}
\includegraphics[width=6cm]{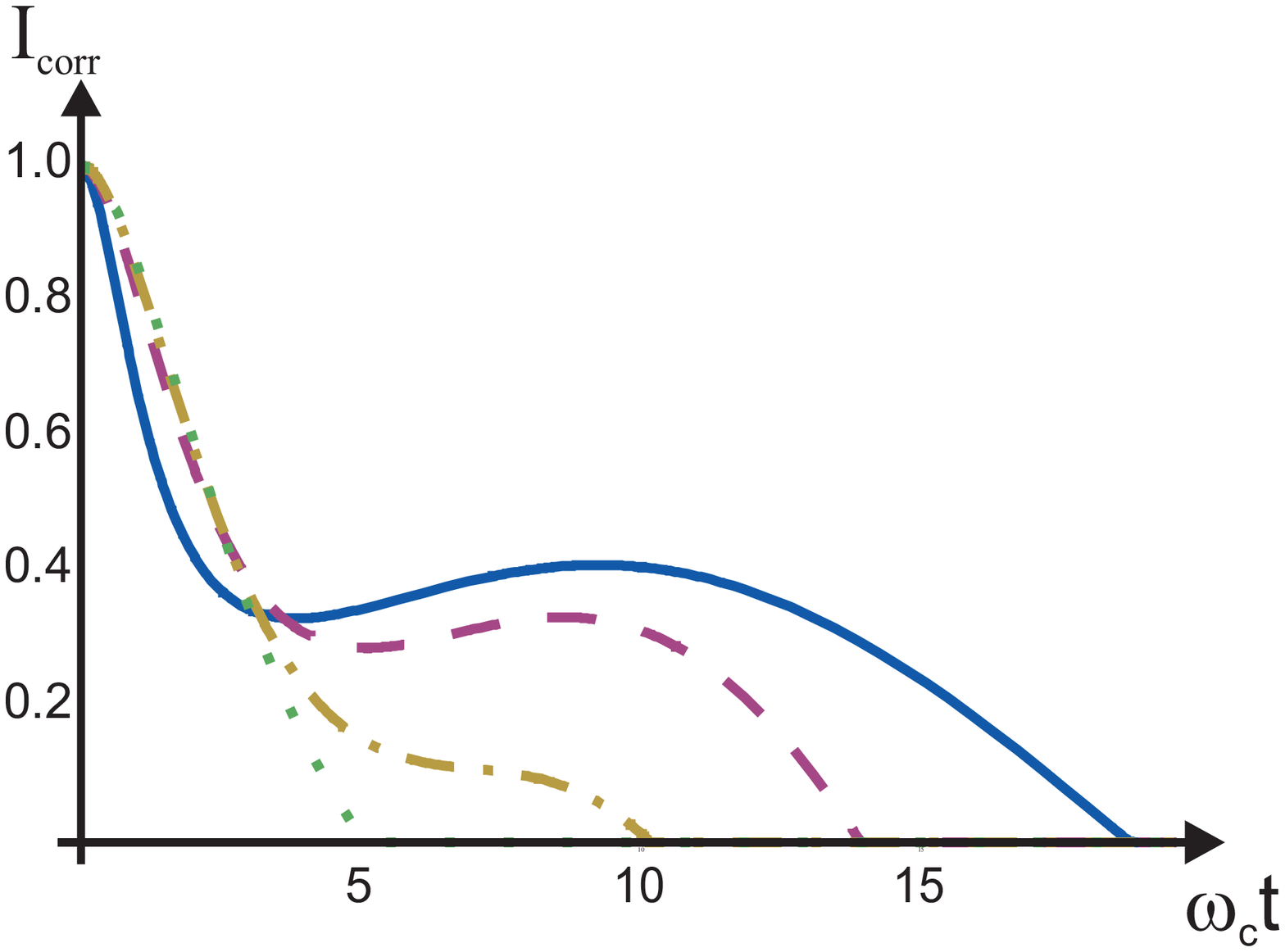}
\includegraphics[width=6cm]{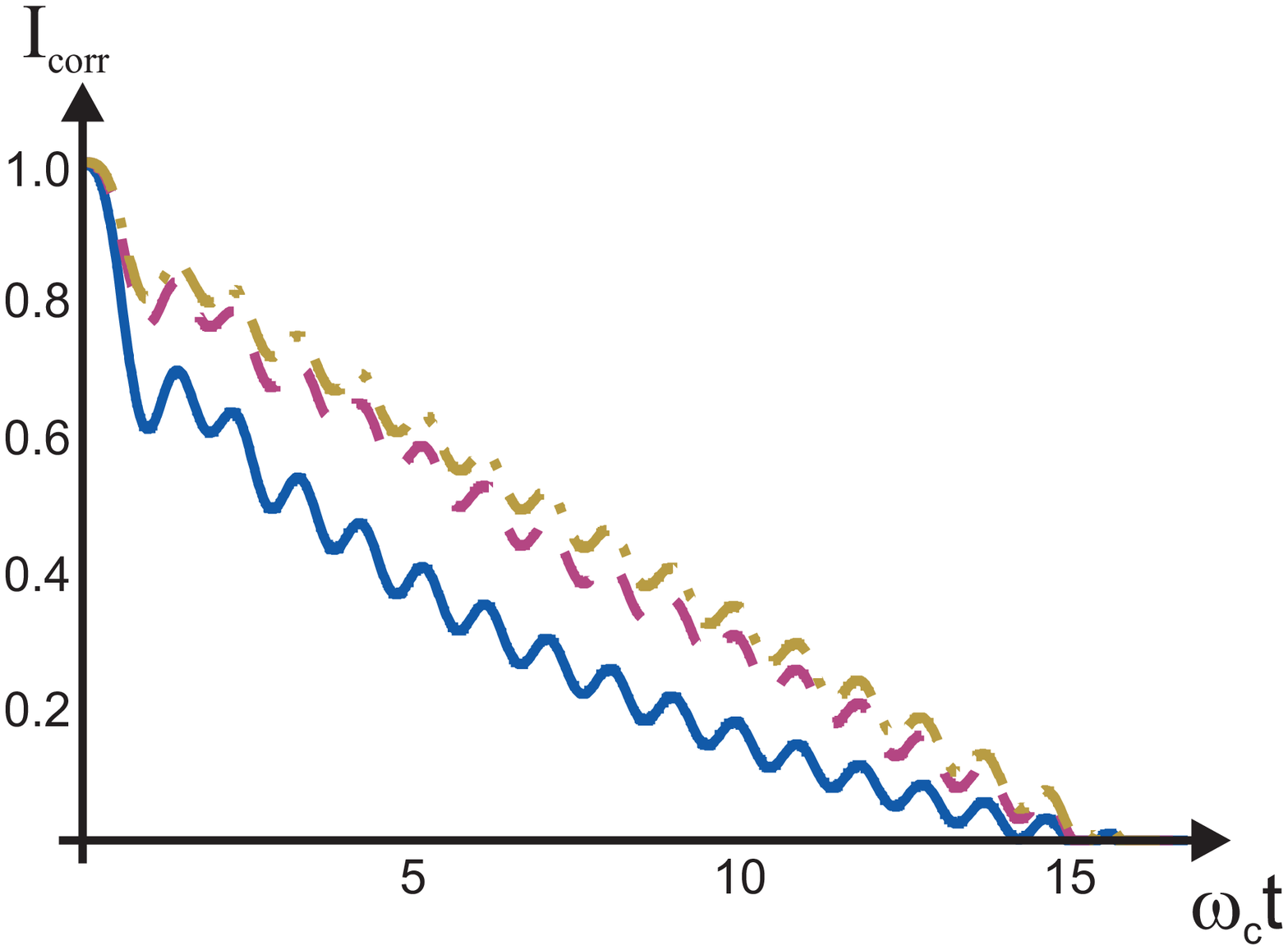}
\end{center}
\vspace{-0.5cm} \caption{(Colors online) Dynamics of the intensity
correlations marker $\mathbf{I}_{\rm corr}$ for a common reservoir as a
function of $\omega_c t$. Parameters: $\alpha=0.1$,
$k_BT/\hbar\omega_c=100$, $N=0$, (top) $x=10$; (bottom) $x=0.3$.
$r=0.5$ (Blue solid line), $r=1$ (Purple dashed line), $r=1.5$
(Yellow dot-dashed line), $r=5$ (Green dotted line).} \label{fig:4}
\end{figure}

Initial states possessing intensity correlations initially below the
SNL, for small values of $N$, always lose them completely in a
finite time. For large $x$ (FIG. \ref{fig:4} (top)) and $N=0$ there
are no revivals as expected in this dynamical regime. A surprising
result is, however, that the larger is the initial two-mode
squeezing the faster the SNL is reached. One would expect, indeed,
that initial states with higher values of intensity correlations
initially below the SNL maintain this quantum property for longer
times than initial states having smaller values of intensity
correlations initially below the SNL. An opposite result is reached
in the case of small $x$ (FIG. \ref{fig:4} (bottom)), where the
environment leads to a faster loss of intensity correlations when
$r$ is smaller. In this case however the SNL limit is reached at the
same time for each value of $r$ and some oscillations and revivals
are present due to the non-secular terms of the solution.

The behavior of entanglement in these systems has been studied in
detail in previous papers \cite{PraBer,PazRon}. We can however
summarize the most important features. In the weak coupling regime,
if there is no initial entanglement, it is not possible to create
it. This has been already pointed out previously for uncorrelated
initial conditions. The same conclusion holds, however, also for
initially correlated states with high enough values of $N$ in the
initial state. In general, for initially entangled states, in the
short-time scale we can observe sudden death and revivals depending
on the values of $N$, $r$ ,$x$. If the initial value of the
entanglement is small, it is rather difficult to observe revivals.
Some revivals can be seen in the $x\ll 1$ case, usually due to
non-secular terms \cite{VasMan}. When the system is initially
strongly correlated entanglement sudden death and revivals can be
observed as well as situations in which in the non-Markovian time
scale the entanglement never goes to zero.

\begin{figure}[h!]
\begin{center}
\includegraphics[width=6cm]{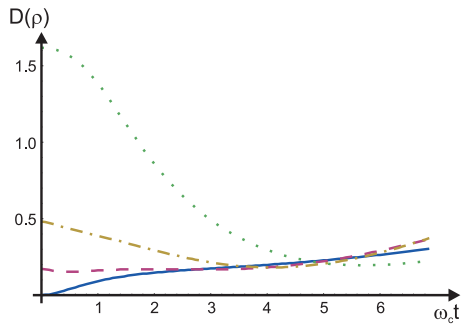}
\end{center}
\vspace{-0.5cm} \caption{(Colors online) Evolution of quantum
discord $\mathcal{D}(\varrho)$ as a function of the scaled time
$\omega_c t $ in the common reservoir scenario. Parameters:
$\alpha=0.1$, $k_BT/\hbar\omega_c=100$, $x=10$, $N=0$. Different
lines represent different values of the squeezing parameter: $r=0$
(Solid blue line), $r=0.2$ (Dashed red line), $r=0.4$ (Dotted-dashed
yellow line) and $r=1$ (Dotted green line).} \label{fig:5}
\end{figure}

Finally, we consider the behavior of the quantum discord. As in the
uncorrelated case, the value of $N$ in the initial state influence
the rate of discord changes. High values of $N$ make the
correlations more robust to the influence of the reservoir. The
value of the resonance parameter $x$ does not influence the
qualitative behavior of quantum discord. Small $x$ lead to a slower
dynamics and presence of oscillations in the solution. So we
consider as an example the case $N=0$ and $x=10$ (See. FIG.
\ref{fig:5} ). The different curves correspond to different values
of $r$. If $r$ is small (also $r=0$) the discord has an initial
small value and starts to increase during the dynamics for short
times. On the contrary if $r$ is large the discord has an initial
high value and it decreases in time. However, independently from the
value of $r$, after an initial transient time interval, all the
discord curves tend to overlap. The interaction with the common bath
seems to destroy the original information on the initial
correlations during such transient time interval, driving the CV
system towards states which have similar value of the quantum
discord.

\section{Conclusions}\label{s:concl}
In this paper we have studied the time evolution of three different
indicators of quantum correlations for a bipartite CV system
initially prepared in a thermal TWB state and interacting with
either independent reservoirs or a common reservoir at high
temperature. Using a definition for the quantum discord in CV
systems recently introduced in Refs \cite{Matt,DaAde}, we have
calculated analytically the dynamics of such quantity under the only
assumption of weak coupling between the system and the environment.
Moreover, we have compared the dynamics of the discord with the
dynamics of entanglement and intensity correlations.

We have demonstrated that, in the independent reservoir scenario,
initially correlated thermal TWB states, more specifically states
with nonzero initial discord, loose their quantum correlations
slower and slower for increasing values of their initial thermal
component $N$. Similarly, in the common reservoir scenario, for
initially uncorrelated states having zero discord, the higher is
$N$, the slower is the reservoir-mediated rate of increase of
quantum discord.

A comparison between the dynamics of discord, entanglement and
intensity correlations shows that, when the two system oscillators
interact with independent reservoirs, initial nonclassical intensity
correlations disappear faster than the initial entanglement, which
in turn disappear faster than the initial discord, for $x\gg 1$. In
the opposite regime $x \ll 1$, the dynamics of the intensity
correlations and of the entanglement is very similar and both
quantities disappear at the same finite time.

In the common reservoir scenario, we have studied whether initial
states possessing zero correlations can become correlated via the
action of the reservoir in the weak coupling case here considered.
Our results show that only discord can be created, while
entanglement and initial correlations remain zero if they are
initially zero. In fact, quantum states with nonzero discord are
much more common than entangled states, as demonstrated, e.g., in
\cite{Ferr}. In this sense we expect quantum discord to be easier to
generate than entanglement.

Our results are the first attempt to characterize the time evolution
of discord and intensity correlations, comparing them to the time
evolution of the entanglement, in CV non-Markovian systems. Hence
they provide a first step in the description of the behavior of
quantum correlations and their robustness under the effect of
dissipative environments. In the future we plan to investigate
whether effects such as the constant discord and the sudden
transition form classical to quantum decoherence \cite{Laura},
recently discovered in bipartite qubits systems, may occur also in
CV systems.

\acknowledgments This work was supported by the Emil Aaltonen
foundation, the Finnish Cultural foundation, the Magnus Ehrnrooth
foundation, by the Turku Collegium of Science and Medicine (S.M.)
and by the CNR-CNISM agreement (S.O.).

\appendix*
\section{The master equation coefficients}

The time-dependent coefficients of the master equations
\eqref{HuPaZang} and \eqref{HuPaZangCom}, in the case of thermal
reservoirs and at the second order in the coupling $\alpha$ are
given by
\begin{subequations}
\begin{align}
\Delta(t)&=\alpha^2\!\!\int_0^t\!\! ds \!
\int_0^{+\infty}\!\!\!\!\!\!\!\! \!d\omega
J(\omega)[2N(\omega)+1]\cos(\omega s)\cos(\omega_0 s), \\
\Pi(t)&=\alpha^2\!\!\int_0^t\!\! ds \!
\int_0^{+\infty}\!\!\!\!\!\!\!\! \!d\omega
J(\omega)[2N(\omega)+1]\cos(\omega s)\sin(\omega_0 s), \\
\gamma(t)&=\alpha^2\!\!\int_0^t\!\! ds \!
\int_0^{+\infty}\!\!\!\!\!\!\!\! \!d\omega
J(\omega)\sin(\omega s)\sin(\omega_0 s), \\
r(t)&=\alpha^2\!\!\int_0^t\!\! ds \!
\int_0^{+\infty}\!\!\!\!\!\!\!\! \!d\omega J(\omega)\sin(\omega
s)\cos(\omega_0 s),
\end{align}
\end{subequations}
with $N(\omega)=[\exp(\hbar\omega/k_BT)-1]^{-1}$ being the mean
number of photons with frequency $\omega$, while $J(\omega)$ defines
the spectral distribution of the environments. For the Ohmic
distribution with Lorentz-Drude cut-off in the high-T limit
($2N(\omega)+1\simeq k_BT/\hbar\omega$) we have
\begin{subequations}\label{Coeff}
\begin{align}
\Delta(t)&=\frac{\alpha^2\omegaz
x^2}{2(1+x^2)}\frac{k_BT}{\hbar\omega_c}
\{x-e^{-\tau}[x\cos(\tau/x)-\sin(\tau/x)\}, \\
\Pi(t)&=\frac{\alpha^2\omegaz
x^2}{2(1+x^2)}\frac{k_BT}{\hbar\omega_c}
\{1-e^{-\tau}[\cos(\tau/x)+x\sin(\tau/x)]\}, \\
\gamma(t)&=\frac{\alpha^2\omegaz x^2}{2(1+x^2)}
\{1-e^{-\tau}[\cos(\tau/x)+x\sin(\tau/x)]\}.
\end{align}
\end{subequations}
We do not provide the analytic expression of $r(t)$ because its
contribution to the solution in the weak coupling limit is
negligible.

\section{The master equation solution}

The time evolution of the characteristic functions in our two
decoherence models \eqref{CharSolInd} and \eqref{CharSolCom} contain
the $2\times 2$ matrices $\mathbf{R}(t)$ and $\bar{\mathbf{W}}(t)$.
Under the weak coupling assumption their expressions is
\begin{equation}
\mathbf{R}(t)=\left(
               \begin{array}{cc}
                 \cos\omegaz t & \sin\omegaz t \\
                 -\sin\omegaz t & \cos\omegaz t \\
               \end{array}
             \right),
\end{equation}
\begin{equation}
\bar{\mathbf{W}}(t)=e^{-\Gamma(t)}\mathbf{R}(t)\biggl[\int_0^te^{-\Gamma(s)}\mathbf{M}(s)ds\biggl]\mathbf{R}^{T}(t),
\end{equation}
with
\begin{equation}
\mathbf{M}(t)=\mathbf{R}^T(t)\left(
               \begin{array}{cc}
                 \Delta(t) & -\Pi(t)/2 \\
                 -\Pi(t)/2 & 0 \\
               \end{array}
             \right)\mathbf{R}(t).
\end{equation}
If we explicit the calculations the following functions appear
\begin{subequations}
\begin{align}
&\Gamma(t)=2\int_0^t\gamma(s) ds, \\
&\Delta_{\Gamma}(t)=e^{-\Gamma(t)}\int_0^te^{\Gamma(s)}\Delta(s) ds, \\
&\Delta_{co}(t)=e^{-\Gamma(t)}\int_0^te^{\Gamma(s)}\Delta(s)\cos[2\omega_0(t-s)]
ds, \\
&\Delta_{si}(t)=e^{-\Gamma(t)}\int_0^te^{\Gamma(s)}\Delta(s)\sin[2\omega_0(t-s)]
ds, \\
&\Pi_{co}(t)=e^{-\Gamma(t)}\int_0^te^{\Gamma(s)}\Pi(s)\cos[2\omega_0(t-s)]
ds, \\
&\Pi_{si}(t)=e^{-\Gamma(t)}\int_0^te^{\Gamma(s)}\Pi(s)\sin[2\omega_0(t-s)]
ds.
\end{align}
\end{subequations}
The last five coefficients can be evaluated numerically. However if
we are interested in the short non-Markovian time scale, we can use
the approximation $\exp(\pm\Gamma(t))\simeq1$. Under this assumption
we can evaluate all the coefficients exactly in the case of the
Lorentz-Drude spectral function at high-T.

The last four coefficients are called non-secular. In some dynamical
regimes their contribution is not essential and can be neglected
(secular approximation). In this paper however we do not perform
this approximation.

\subsection{Independent reservoir solution}
As we already pointed out, for an initial Gaussian state with zero
mean, the solution of the master equation is obtained giving the
time evolution of the covariance matrix. In the case of independent
reservoirs we need to apply definitions \eqref{CharSolInd} and
\eqref{Charfun} to Eq. \eqref{CovMatInd}. If we do this we find that
the covariance matrix \eqref{CovMat} at time $t$ can be written as
\begin{subequations}
\begin{align}
&\mathbf{A^{ind}_t}=\mathbf{A}_0e^{-\Gamma}+
\Delta_{\Gamma}{\mathbbm1} + \left(\begin{array}{cc}
(\Delta_{co}-\Pi_{si}) & -\Delta_{si}+\Pi_{co} \\
-\Delta_{si}+\Pi_{co} & -(\Delta_{co}-\Pi_{si}) \\
\end{array}\right),\\
&\mathbf{C^{ind}_t}=\left(N+\frac12\right)\sinh(2r)\,e^{-\Gamma}
\left( \begin{array}{cc}
\cos2\omegaz t &  \sin2\omegaz t \\
\sin2\omegaz t & -\cos2\omegaz t \\
\end{array}\right).
\end{align}
\end{subequations}

\subsection{Common reservoir solution}

To obtain the solution for the common reservoir model we first have
to transform the original covariance matrix $\sigma(0)$ into the one
in the new picture $\tilde{\sigma}(0)$. Then using Eqs.
\eqref{CharSolCom} and \eqref{CovMatComm} we evolve the matrix into
$\tilde{\sigma}(t)$ and finally we apply the inverse picture
transformation to get the solution to the problem $\sigma(t)$. The
transformation expressions can be easily obtained by comparing the
following definitions of the covariance matrices and implementing
the canonical relations
\begin{subequations}
\begin{align}
&\sigma_{ij}=\langle\{\hat{\Lambda}_i,\hat{\Lambda}_j\}\rangle/2-\langle\hat{\Lambda}_i\rangle\langle\hat{\Lambda}_j\rangle,\\
&\tilde{\sigma}_{ij}=\langle\{\hat{\Lambda}^{\pm}_i,\hat{\Lambda}^{\pm}_j\}\rangle/2-\langle\hat{\Lambda}^{\pm}_i\rangle\langle\hat{\Lambda}^{\pm}_j\rangle,
\end{align}
\end{subequations}
with $\hat{\Lambda}\equiv(\hat{X}_1,\hat{P}_1,\hat{X}_2,\hat{P}_2)$
and
$\hat{\Lambda^{\pm}}\equiv(\hat{X}_+,\hat{P}_+,\hat{X}_-,\hat{P}_-)$.
The solution then reads
\begin{equation}
\sigma(t)=\left(
            \begin{array}{cccc}
              \chi & z & \mu & \xi \\
              z & y & \xi & \nu \\
              \mu & \xi & \chi & z \\
              \xi & \nu & z & y \\
            \end{array}
          \right)
\end{equation}
where

\begin{subequations}
\begin{align}
&\chi=g_{+}(\Gamma)\,a - g_{-}(\Gamma)\,c\,\cos(2x)
+\frac{\Delta_{\Gamma}+(\Delta_{co}-\Pi_{si})}{\sqrt{2}}\\
&y=g_{+}(\Gamma)\,a + g_{-}(\Gamma)\,c\,\cos(2x)
+\frac{\Delta_{\Gamma}-(\Delta_{co}-\Pi_{si})}{\sqrt{2}}\\
&z=g_{-}(\Gamma)\,c\,\sin(2x)-\frac{\Delta_{si}-\Pi_{co}}{\sqrt{2}}
\end{align}
\end{subequations}
and
\begin{subequations}
\begin{align}
&\mu=-g_{-}(\Gamma)\,a + g_{+}(\Gamma)\,c\,\cos(2x)
+\frac{\Delta_{\Gamma}+(\Delta_{co}-\Pi_{si})}{\sqrt{2}}\\
&\nu=-g_{-}(\Gamma)\,a - g_{+}(\Gamma)\,c\,\cos(2x)
+\frac{\Delta_{\Gamma}-(\Delta_{co}-\Pi_{si})}{\sqrt{2}}\\
&\xi=-g_{+}(\Gamma)\,c\,\sin(2x)-\frac{\Delta_{si}-\Pi_{co}}{\sqrt{2}}
\end{align}
\end{subequations}
with $g_{\pm}(\Gamma)=\frac12 (1\pm e^{-\Gamma})$,
$a=(N+1/2)\cosh(2r)$ and $c=(N+1/2)\sinh(2r)$.


\begin{thebibliography}{30}

\bibitem{Wer}
R. F. Werner, Phys. Rev. A \textbf{40}, 4277-4281 (1989).
\bibitem{OllZur}
H. Ollivier and W. H. Zurek, Phys. Rev. Lett. \textbf{88}, 017901
(2001).
\bibitem{HenVed}
L. Henderson and V. Vedral, J. Phys. A \textbf{34}, 6899 (2001).
\bibitem{DatCav}
A. Datta, A. Shaji, and C. M. Caves, Phys. Rev. Lett. \textbf{100},
050502 (2008).
\bibitem{LanWhi}
B. P. Lanyon, M. Barbieri, M. P. Almeida, and A. G. White, Phys.
Rev. Lett. \textbf{101}, 200501 (2008).
\bibitem{Lid09} A. Shabani, D. A. Lidar, Phys. Rev. Lett. {\bf 102}
100402 (2009).
\bibitem{Luo} S. Luo, Phys. Rev. A \textbf{77}, 042303 (2008).
\bibitem{Ali}
M. Ali, A. R. P. Rau, and G. Alber, Phys. Rev. A \textbf{81}, 042105
(2010).
\bibitem{MarDisc}
J. Maziero, L. C. Celeri, R. M. Serra, and V. Vedral, Phys. Rev. A
\textbf{80}, 044102 (2009); J. Maziero, T. Werlang, F.F. Fanchini,
L. C. Celeri, and R. M. Serra, Phys. Rev. A \textbf{81}, 022116
(2010).
\bibitem{NMDisc1}
T. Werlang, S. Souza, F. F. Fanchini, and C.J. Villas-Boas, Phys.
Rev. A \textbf{80}, 024103 (2009).
\bibitem{NMDisc2}
F. F. Fanchini, T. Werlang, C. A. Brasil, L. G. E. Arruda, and A. O.
Caldeira, arXiv:0911.1096.
\bibitem{NMDisc3}
Bo Wang, Zhen-Yu Xu, Ze-Qian Chen, and Mang Feng, Phys. Rev. A
\textbf{81}, 014101 (2010).
\bibitem{YuEbe}
Ting Yu and J. H. Eberly Science \textbf{30} January 2009 323:
598-601.
\bibitem{Laura}
L. Mazzola, J. Piilo, and S. Maniscalco, arXiv:1001.5441.
\bibitem{NatComm}
Jin-Shi Xu \emph{et al.}, Nat. Commun. \textbf{1}, 7 (2010).
\bibitem{Matt}
P. Giorda and M. G. A. Paris, arXiv:1003.3207.
\bibitem{DaAde}
A. Datta and G. Adesso, arXiv:1003.4979.
\bibitem{DBA07} I. P. Degiovanni, M. Bondani, E. Puddu, A. Andreoni,
M. G. A. Paris, Phys. Rev. A \textbf{76}, 062309 (2007).
\bibitem{ss07} M. Bondani, A. Allevi, G. Zambra, M. G. A. Paris, A.
Andreoni, Phys. Rev. A {\bf 76} 013833 (2007).
\bibitem{ThsPDC}
I. P Degiovanni, M. Genovese, V. Schettini, M. Bondani, A. Andreoni,
M G A Paris, Phys. Rev. A {\bf 79}, 063836 (2009).
\bibitem{SNL}
E. N. Gilbert and H. O. Pollak, Bell Syst. Tech. J. \textbf{39}, 333
(1960); C. M. Caves, Phys. Rev. D \textbf{26}, 1817 (1982); H. P.
Yuen and V. W. S. Chan, Opt. Lett. \textbf{8}, 177 (1983).
\bibitem{PraBer}
J. S. Prauzner-Bechcicki, J. Phys. A \textbf{37}, L173 (2004).
\bibitem{PazRon}
J. P. Paz and A. J. Roncaglia, Phys. Rev. Lett. \textbf{100}, 220401
(2008); J. P. Paz and A. J. Roncaglia, Phys. Rev. A \textbf{79},
032102 (2009).
\bibitem{IndepRes} M. Ban, J. Phys. A {\bf 39}, 1927 (2006); M. Ban,
  Phys. Lett. A {\bf 359}, 402 (2006); Jun-Hong An and Wei-Min Zhang,
  Phys. Rev. A {\bf 76} 042127 (2007); Kuan-Liang Liu and Hsi-Sheng
  Goan, Phys. Rev. A {\bf 76}, 022312 (2007); Jun-Hong An, Ye Yeo,
  Wei-Min Zhang, and C. H. Oh, J. Phys.  A: Math. Theor. {\bf 42},
  015302 (2009); K. Shiokawa, Phys. Rev. A {\bf 79}, 012308 (2009).
\bibitem{VasMan}
S. Maniscalco, S. Olivares, and M. G. A. Paris, Phys. Rev. A
\textbf{75}, 062119 (2007); R. Vasile, S. Olivares, M. G. A. Paris,
and S. Maniscalco, Phys. Rev. A \textbf{80}, 062324 (2009).
\bibitem{HuPaZa}
B. L. Hu, J. P. Paz, and Y. Zhang, Phys. Rev. D \textbf{45}, 2843
(1992).
\bibitem{Intra} F. Intravaia, S. Maniscalco, and A.
Messina, Phys. Rev. A \textbf{67}, 042108 (2003).
\bibitem{Icorr:brida} G. Brida, M. Bondani, I. P. Degiovanni,
  M. Genovese, M. G. A. Paris, I. Ruo Berchera and V. Schettini , e-print
  arXiv:0909.5288 [quant-ph].
\bibitem{ferr:prep} A.~Ferraro and M.~G.~A.~Paris, in preparation.
\bibitem{Simon}
R. Simon, 2000 Phys. Rev. Lett. \textbf{84}, 2726 (2000).
\bibitem{EntFor}
G. Giedke, M. M. Wolf, O. Kruger, R. F. Werner, and J. I. Cirac, Phys.
Rev. Lett. 91, 107901 (2003).
\bibitem{buono:JOSAB:10} D. Buono, G. Nocerino, V. D'Auria, A. Porzio,
S. Olivares and M. G. A. Paris, J. Opt. Soc. Am. B {\bf 27}, A110
(2010).
\bibitem{Nega}
G. Vidal, and R. F. Werner, Phys. Rev. A \textbf{65}, 032314 (2002).
\bibitem{Ferr}
A. Ferraro, L. Aolita, D. Cavalcanti, F. M. Cucchietti, A. Acin
arXiv:0908.3157 (2009).


\end{thebibliography}
\end{document}